\documentclass[12pt,showpacs,preprintnumbers,amsmath,amssymb]{revtex4}
\usepackage{graphicx}
\begin{document}

\newcommand{\pderiv}[2]{\frac{\partial #1}{\partial #2}}
\newcommand{\deriv}[2]{\frac{d #1}{d #2}}

\title{Destruction of First-Order Phase Transition in a Random-Field Ising
Model}

\vskip \baselineskip

\author{Nuno Crokidakis}
\thanks{E-mail address: nuno@if.uff.br}

\author{Fernando D. Nobre}
\thanks{Corresponding author: E-mail address: fdnobre@cbpf.br}

\address{
Centro Brasileiro de Pesquisas F\'{\i}sicas \\
Rua Xavier Sigaud 150 \\
22290-180 \hspace{5mm} Rio de Janeiro - RJ \hspace{5mm} Brazil}

\date{\today}


\begin{abstract}
\noindent
The phase transitions that occur in an infinite-range-interaction Ising
ferromagnet in the presence  
of a double-Gaussian random magnetic field are analyzed. 
Such random fields are defined as a superposition of two Gaussian 
distributions, presenting
the same width $\sigma$. It is argued that this distribution is more 
appropriate for a theoretical description of real systems than its simpler
particular 
cases, i.e., the bimodal ($\sigma=0$) and single Gaussian distributions.
It is shown that a 
low-temperature first-order phase transition may be destructed for increasing
values of $\sigma$, similarly to what happens in the compound 
${\rm Fe_{x}Mg_{1-x}Cl_{2}}$, whose finite-temperature first-order phase
transition is presumably destructed by an increase in the field
randomness.  

\vskip \baselineskip


\end{abstract}
\maketitle

\newpage

\noindent
\section{Introduction}


The theory of continuous phase transitions is very well established
nowadays \cite{stanleybook,yeomansbook,cardybook}; these transitions are
characterized by a divergence of the correlation length, in such a way that
some microscopic 
details of the system become irrelevant, leading to the concept
of universality classes for the critical exponents. Although first-order
phase transitions are very common in nature, they have attracted much less
attention from the theoretical point of view; in particular, 
the correlation length remains finite at such transitions, and so a
universal behavior is not expected. 
Many interesting features occur in first-order phase transitions, like a
discontinuity in the order parameter and the presence of latent heat; in
addition to those, in some systems, 
it is possible for a first-order phase transition to become
continuous due to changes in an external parameter, or to the presence
of microscopic inhomogeneities that may generate significant roundings in
the transition \cite{imrywortis}. 
Therefore, one important question
concerns the effects of random quenched impurities on the
thermodynamic phase transition. 

The random-field Ising model (RFIM) was introduced by Imry and Ma 
\cite{imryma} and has attracted
a lot of interest after the identification of its physical
realizations. Probably the most important physical conception of the RFIM
comes out to be a diluted Ising antiferromagnet in the presence of a uniform
magnetic field \cite{fishmanaharony,cardy}. In these systems one has local
variations in the sum of exchange couplings that connect a given site to
other sites of the system, leading to local variations of the
two-sublattice site magnetizations, and as a consequence, one may have local
magnetizations that vary in both sign and magnitude. In these
identifications, the effective random field at a given site is expressed
either in terms of its local magnetization \cite{fishmanaharony}, or of the
sum of the exchange couplings associated to this site \cite{cardy}. 

Since then, many diluted
antiferromagnets have been investigated, in such a way that systems like
${\rm Fe_{x}Zn_{1-x}F_{2}}$ and ${\rm Fe_{x}Mg_{1-x}Cl_{2}}$, for certain
ranges of the concentration x (essentially high values of x, as mentioned
above) are nowadays 
considered as standard experimental realizations of the RFIM
\cite{belangerreview}. These
systems are characterized by large crystal-field anisotropies, so
that they may be reasonably well-described in terms of Ising variables.  
In particular, the compound ${\rm Fe_{x}Mg_{1-x}Cl_{2}}$ behaves like an
Ising spin glass for ${\rm x} < 0.55$, and is considered as a typical RFIM
for higher magnetic concentrations. In the RFIM regime, it presents a
very curious behavior: one finds a
first-order transition turning into a continuous one 
due to a change in the random fields
\cite{belangerreview,kushauerkleemann,kushauer}; the concentration at which
the first-order transition disappears is estimated to be $x=0.6$. 

From the theoretical point of view, many
important questions remain open. 
At the mean-field level, it is well
known that different probability distributions for the random fields may
lead to distinct phase diagrams, e.g., a Gaussian probability
distribution yields a continuous ferromagnetic-paramagnetic boundary
\cite{schneiderpytte}, whereas for a bimodal distribution, this boundary
exhibits a continuous piece at high temperatures ending up at a tricritical
point, followed by a first-order phase transition at low
temperatures \cite{aharony78}. Indeed, it was argued  that
whenever an analytic symmetric distribution for the fields presents a minimum
at zero field, one should expect a tricritical point and a first-order
transition 
for sufficiently low temperatures \cite{aharony78}; such an argument has
been considered further and improved by other authors
\cite{andelman,galambirman}. 
The presence of such a first-order transition 
for sufficiently low temperatures in short-range-interaction models,
represents a point that has not been fully elucidated.
For the three-dimensional RFIM, high-temperature series expansions
\cite{gofman} and a zero-temperature scaling analysis \cite{swift} find
continuous transitions for both Gaussian and bimodal distributions.
Nevertheless,
several zero-temperature studies of the Gaussian three-dimensional RFIM
\cite{machta00,middleton,machta03}
suggest a first-order transition for such a model; this occurs because the
magnetization 
and specific-heat critical exponents are very small, and so, it is
difficult to determine whether the magnetization vanishes continuously, or
discontinuously, at the transition.
However, in four dimensions
a zero-temperature analysis \cite{swift} leads to a first-order
transition in the bimodal case and a continuous one for a Gaussian
distribution, in agreement with the mean-field predictions.
 
The crossover from first-order to continuous phase
transitions has been investigated through different
theoretical approaches \cite{kushauer,aizenman,huiberker,sethna93}. One
possible mechanism used to find such a crossover, or even to suppress the 
first-order
transition completely, consists in introducing an additional kind of randomness in
the system, e.g., bond randomness \cite{aizenman,huiberker}. By considering randomness in the field only, this crossover
has been also analyzed 
through zero-temperature studies, either within mean-field theory
\cite{sethna93}, or numerical simulations on a three-dimensional lattice
\cite{kushauer,sethna93}. Since the effects observed in the diluted
antiferromagnet ${\rm Fe_{x}Mg_{1-x}Cl_{2}}$ occur for rather low
temperatures (typically around 4 K), they may be described satisfactorily by
zero-temperature approaches. However, such zero-temperature approaches
would not be appropriate for explaining similar effects at higher
temperatures, that may possibly be found on similar systems.

In the present work we present a RFIM that is able to exhibit a
first-order phase 
transition turning into a continuous one, and characterized by a
destruction of 
the first-order phase transition due a change in the random fields, at
finite temperatures. The interactions among spins are of infinite range,
i.e., in the limit where the mean-field approach is exact, and the random
fields are defined through a double 
Gaussian probability distribution, which recovers in certain limits, the
simple Gaussian and bimodal probability distributions. In the next section
we present the model, find its free energy and phase diagrams. In section 3
we present our conclusions. 

\noindent
\section{The Model and Its Solution}

 
Let us define the infinite-range-interaction Ising model in the
presence of an external  
random magnetic field, in terms of the Hamiltonian

\begin{equation} \label{1}
\mathcal{H}=- \frac{J}{N}\sum_{(i,j)}S_{i}S_{j} - \sum_{i}H_{i}S_{i}~, 
\end{equation}

\vskip \baselineskip
\noindent
where $S_{i}=\pm 1$ ($i=1,2,...,N$) and the sum $\sum_{(i,j)}$ applies to
all distinct pairs of spins. The random fields $\{H_{i}\}$ are  quenched
variables, following a double Gaussian probability distribution, 

\begin{equation} \label{2}
P(H_{i})=\frac{1}{2}\left(\frac{1}{2\pi \sigma^{2}}\right)^{1/2}
\left\{\exp\left[-\frac{(H_{i}-H_{0})^{2}}{2\sigma^{2}}\right]
+\exp\left[-\frac{(H_{i}+H_{0})^{2}}{2\sigma^{2}}\right]\right\}~. 
\end{equation}

\begin{figure}[t]
\begin{center}
\includegraphics[width=0.40\textwidth,angle=0]{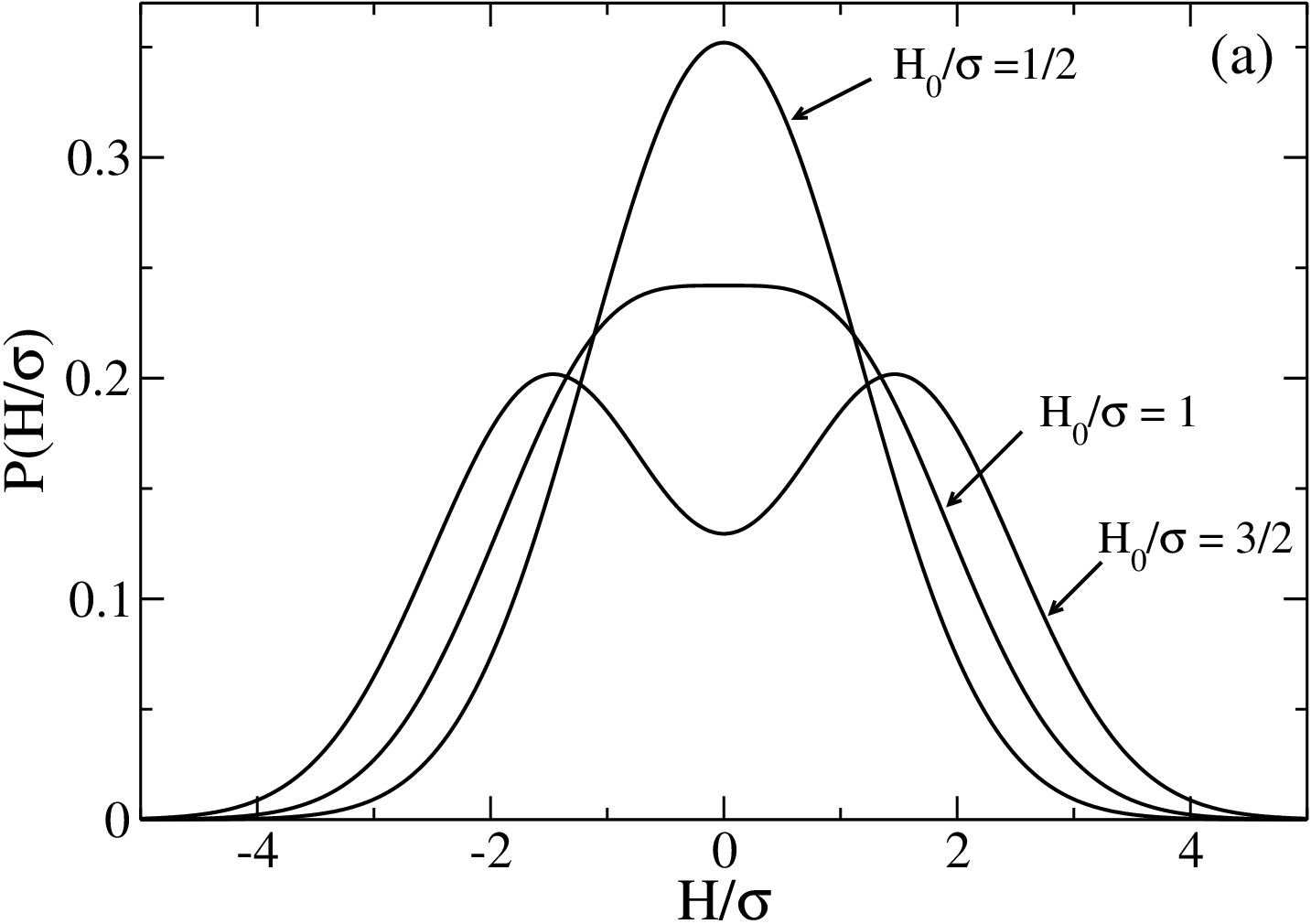}
\hspace{1.5cm}
\includegraphics[width=0.40\textwidth,angle=0]{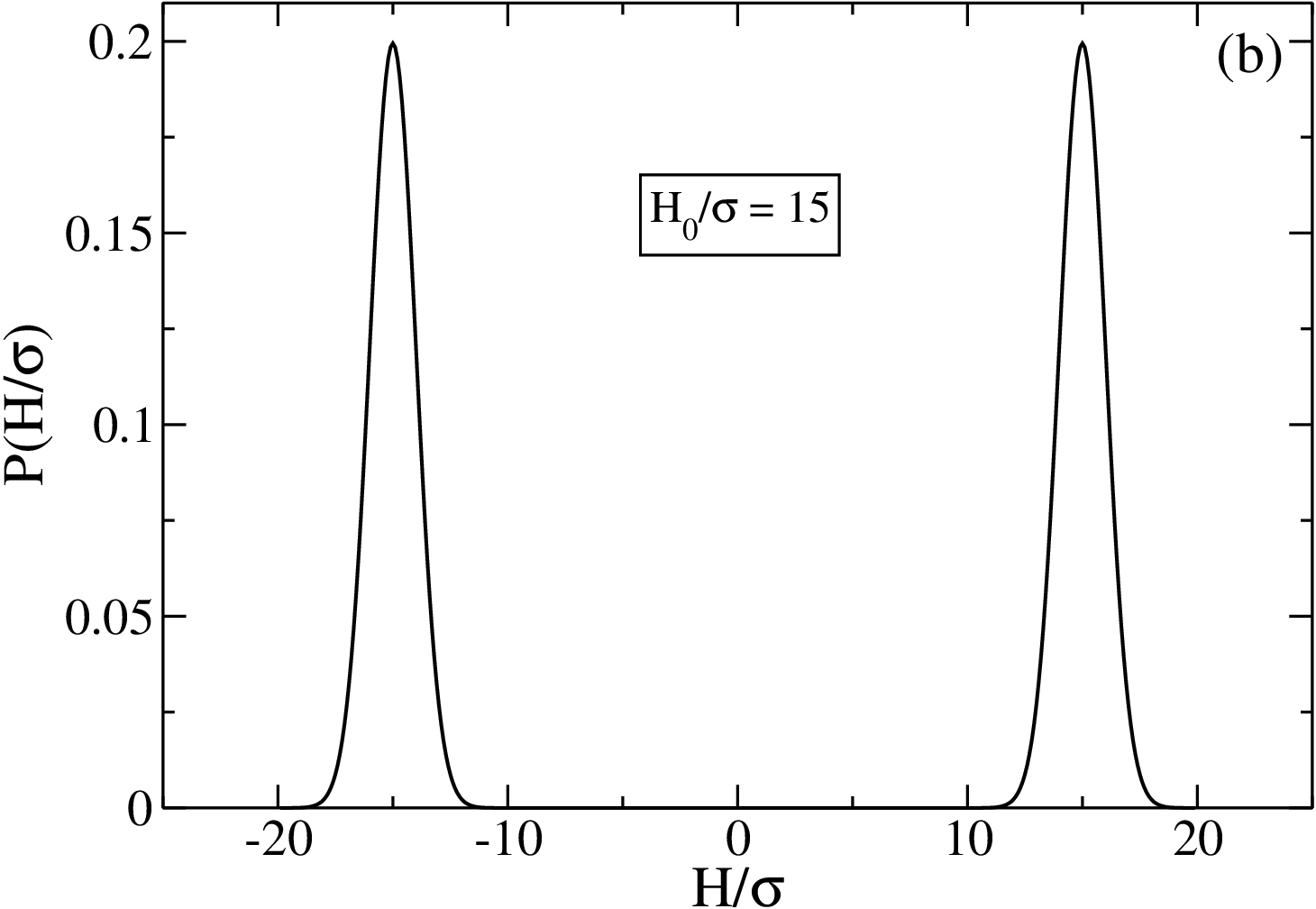}
\end{center}
\protect\caption{The probability distribution of Eq.~(\ref{2}) for
typical values of the ratio $H_{0}/\sigma$: (a) $H_{0}/\sigma=1/2,1,3/2$; 
(b) $H_{0}/\sigma=15$.}  
\label{fig1}
\end{figure}

\vskip \baselineskip
\noindent
The probability distribution above depends on two parameters, $H_{0}$ and
$\sigma$, and modifies its form according to the ratio $H_{0}/\sigma$, as
exhibited in Fig.~1. Such a distribution is
double-peaked for $H_{0}/\sigma>1$, presents a single peak for $H_{0}/\sigma<1$, changing its concavity at the
origin when $H_{0}/\sigma=1$. Besides that, in the limits $H_{0} \rightarrow 0$ and 
$\sigma \rightarrow 0$, one recovers the symmetric 
Gaussian and bimodal probability distributions, respectively. Changes between 
these two limits may be followed by analyzing the moments 
$<H_{i}^{n}>_{H}$ ($n=1,2,3, \cdots$) and in particular, through the kurtosis, 
$\kappa =  <H_{i}^{4}>_{H}/[3(<H_{i}^{2}>_{H})^{2}]$, that varies from
$\kappa=1/3$ (bimodal limit) up to $\kappa=1$ (Gaussian limit), approaching
unit in the limit $H_{0}/\sigma \rightarrow 0$, in which case one gets
a perfect Gaussian distribution. For finite values of $H_{0}/\sigma$ one gets
$1/3<\kappa<1$, and in particular, for the cases exhibited in Fig.~1 one has
that $\kappa \approx 0.97 \ (H_{0}/\sigma=1/2)$, 
$\kappa \approx 0.83 \ (H_{0}/\sigma=1)$, 
$\kappa \approx 0.68 \ (H_{0}/\sigma=3/2)$, 
and $\kappa \approx 0.34 \ (H_{0}/\sigma=15)$.  

For a given realization of the site fields $\{H_{i}\}$, one has a
corresponding free energy $F(\{H_{i}\})$, in such a way that the average 
over disorder, $[F(\{H_{i}\})]_{H}$, becomes 

\begin{equation}  \label{3}
[F(\{H_{i}\})]_{H}=\int\prod_{i}[dH_{i}P(H_{i})]F(\{H_{i}\})~. 
\end{equation}

\vskip \baselineskip
\noindent
One can now make use of the replica method 
\cite{dotsenkobook,nishimoribook} in order to get
the free energy per spin as 

\begin{equation} \label{4}
-\beta f=\lim_{N \to \infty}\frac{1}{N}[\ln Z(\{H_{i}\})]_{H} 
= \lim_{N \to \infty}\lim_{n \to 0}\frac{1}{Nn}([Z^{n}]_{H}-1)~,   
\end{equation}

\vskip \baselineskip
\noindent
where $Z^{n}$ is the partition function of $n$ copies of the original
system defined in Eq.~(\ref{1}) and $\beta = 1/(kT)$. Standart
calculations lead to 

\begin{equation}  \label{5}
\beta f= \lim_{n \to 0} \frac{1}{n} \, {\rm min} \; g(m^{\alpha})~, 
\end{equation}

\vskip \baselineskip
\noindent
where

\begin{equation}  \label{6}
g(m^{\alpha})=\frac{\beta J}{2}\sum_{\alpha}(m^{\alpha})^{2}
-\frac{1}{2}\ln{\rm Tr}_{\alpha}\exp(\mathcal{H}_{{\rm eff}}^{+}) 
-\frac{1}{2}\ln{\rm Tr}_{\alpha}\exp(\mathcal{H}_{{\rm eff}}^{-})~,   
\end{equation}

\begin{equation}  \label{7}
\mathcal{H}_{{\rm eff}}^{\pm}=\beta J\sum_{\alpha}m^{\alpha}S^{\alpha} 
+ \beta \sigma \left( \sum_{\alpha}S^{\alpha} \right)^{2}
\pm \beta H_{0}\sum_{\alpha}S^{\alpha}~.   
\end{equation}

\vskip \baselineskip
\noindent
In the equations above, the index $\alpha$ ($\alpha=1,2,...,n$) is a
replica label and ${\rm Tr}_{\alpha}$ represents a trace over the spin
variables of each replica.
The extrema of the functional $g(m^{\alpha})$ yields the equilibrium
equation for the magnetization of replica $\alpha$, 

\begin{equation} \label{8}
m^{\alpha}=\frac{1}{2}<S^{\alpha}>_{+}+\frac{1}{2}<S^{\alpha}>_{-}~,   
\end{equation}

\vskip \baselineskip
\noindent
where $<\,>_{\pm}$ refer to thermal averages with respect to the
``effective Hamiltonians'' $\mathcal{H}_{{\rm eff}}^{\pm}$ in 
Eq.~(\ref{7}). 

If one assumes the replica-symmetry ansatz 
\cite{dotsenkobook,nishimoribook}, i.e., 
$m^{\alpha}=m$ ($\forall~\alpha$),
the free energy per spin [cf. Eqs.~(\ref{5})--(\ref{7})] and the
equilibrium condition, Eq.~(\ref{8}), become 

\begin{eqnarray} \nonumber
f & = & \frac{J}{2}m^{2} - \frac{1}{2\beta}\frac{1}{\sqrt{2\pi}}
\int_{-\infty}^{+\infty}dze^{-z^{2}/2}\ln2\cosh\Phi^{+} \\ \label{9}
& - & \frac{1}{2\beta}\frac{1}{\sqrt{2\pi}}
\int_{-\infty}^{+\infty}dze^{-z^{2}/2}\ln2\cosh\Phi^{-}~,  \\ \nonumber 
\\ \label{10}
m & = & \frac{1}{2}\frac{1}{\sqrt{2\pi}}
\int_{-\infty}^{+\infty}dze^{-z^{2}/2}\tanh\Phi^{+} 
+ \frac{1}{2}\frac{1}{\sqrt{2\pi}}
\int_{-\infty}^{+\infty}dze^{-z^{2}/2}\tanh\Phi^{-}~,
\end{eqnarray}

\vskip \baselineskip
\noindent
where $\Phi^{\pm}=\beta (Jm + \sigma z \pm H_{0})$.

It is important to mention that the present system exhibits no
instability associated with the replica-symmetric solution \cite{at}, which
usually appears due to parameters characterized by two replica indices,
like in the spin-glass problem \cite{dotsenkobook,nishimoribook}. 
In the RFIM, one has a single phase transition associated with the
magnetization and two phases are possible, namely, the
ferromagnetic ($m \neq 0$) and the paramagnetic ($m = 0$) ones. The
critical frontier separating these two phases may be found by solving 
Eq.~(\ref{10}); in the case of first-order phase transitions, we shall
make use of the free energy per spin, Eq.~(\ref{9}), as well. 
Let us then expand Eq.~(\ref{10}) in powers of $m$, 

\begin{equation} \label{11}
m = Am + Bm^{3} + Cm^{5} + O(m^{7})~, 
\end{equation}

\vskip \baselineskip
\noindent
where the coefficients are given by

\begin{eqnarray} \label{12}
A &=& \beta J\{1-\rho_{1}\}~, \\  \label{13}
B &=& -\frac{(\beta J)^{3}}{3}\{1-4\rho_{1}+3\rho_{2}\}~,  \\ \label{14}
C &=& \frac{(\beta J)^{5}}{15}\{2-17\rho_{1}+30\rho_{2}-15\rho_{3}\}~,
\end{eqnarray}

\vskip \baselineskip
\noindent
with

\begin{equation} \label{15}
\rho_{k}=\frac{1}{\sqrt{2\pi}}\int_{-\infty}^{+\infty}dze^{-z^{2}/2}
\tanh^{2k}\beta(H_{0}+\sigma z)~.  
\end{equation}

\vskip \baselineskip
\noindent
The continuous critical frontier is determined by setting $A=1$,
provided that $B<0$. If a first-order critical
frontier also occurs, the continuous line ends when $B=0$;  
in such cases, the continuous and first-order critical frontiers meet at a
tricritical point, whose coordinates may be obtained by solving the equations 
$A=1$ and $B=0$ numerically. In addition to that, for $B>0$, the
first-order critical frontier may be found by equating the free energies at
each side of this line, i.e.,  
$f(m=0)=f(m\ne0)$.

\begin{figure}[t]
\begin{center}
\includegraphics[width=0.60\textwidth,angle=0]{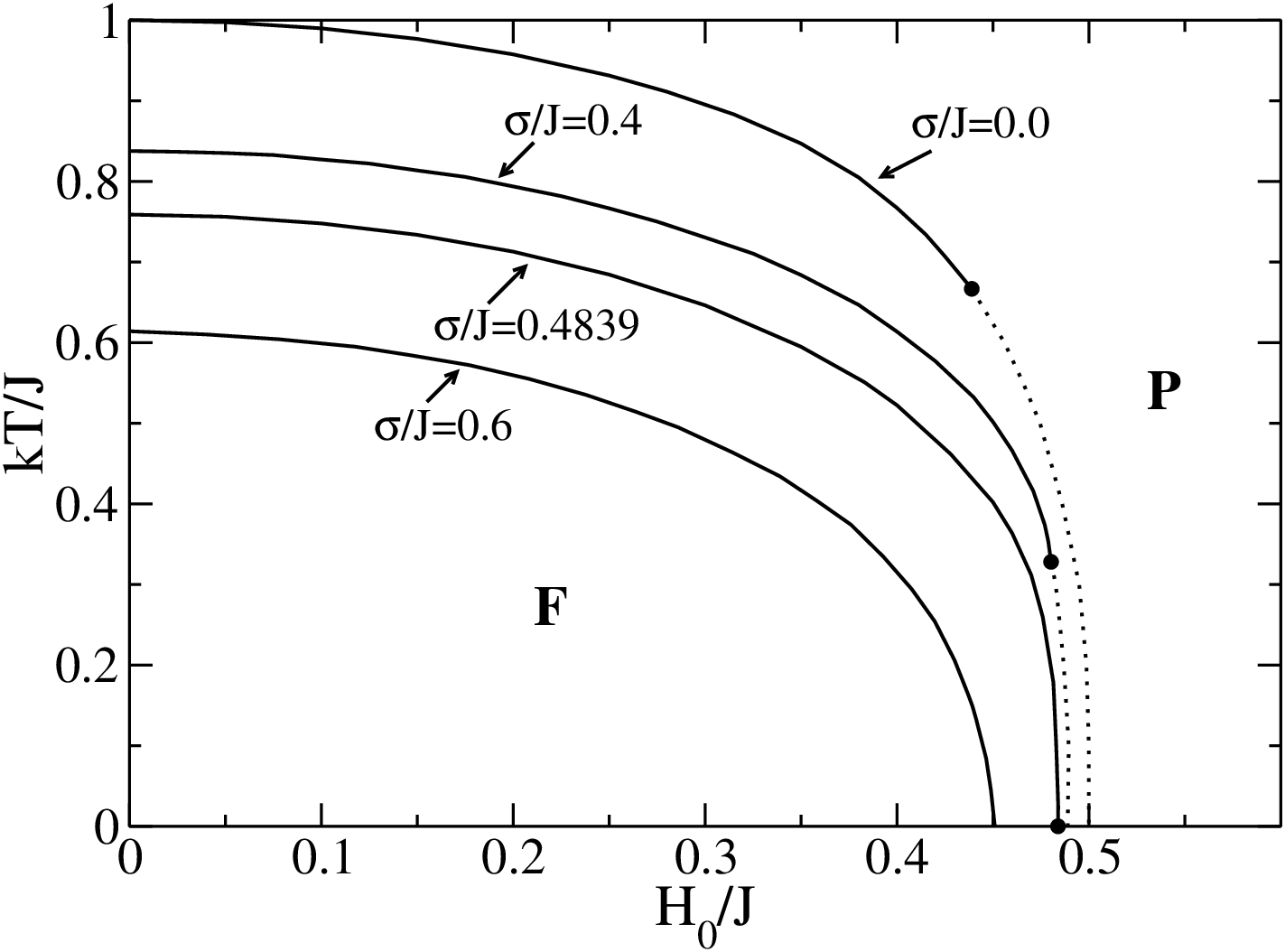}
\end{center}
\protect\caption{Critical frontiers separating the paramagnetic ({\bf P})
and ferromagnetic ({\bf F}) phases of the infinite-range-interaction
ferromagnet in the presence of
a double Gaussian random field, for typical values of $\sigma /J$. Full
(dashed) lines represent continuous (first-order) phase transitions; 
the black circles are tricritical points. Notice the colapse of the
tricritical point at the zero-temperature axis 
for $\sigma/J=0.4839$.}
\label{fig2}
\end{figure}
 
Using this procedure, we have calculated numerically 
the critical frontiers separating the paramagnetic and 
ferromagnetic phases for typical values of $\sigma/J$ (see Fig.~2). As will
be seen below,  
the existence of a tricritical point at finite temperatures is 
restricted to the condition   
$0 \leq (\sigma/J) \leq \sqrt{2/e\pi}$. At the threshold value 
$(\sigma/J) = \sqrt{2/e\pi} \cong 0.4839$ the tricritical point occurs at
zero temperature. 

Let us now analyze how the above-mentioned critical frontier evolves
along the zero-temperature axis; at $T=0$, the free energy and
magnetization become, respectively, 

\begin{eqnarray}\nonumber
f&=&-\frac{J}{2}m^{2}-\frac{H_{0}}{2}
\left[{\rm erf}\left(\frac{Jm+H_{0}}{\sigma\sqrt{2}}\right) 
- {\rm erf}\left(\frac{Jm-H_{0}}{\sigma\sqrt{2}}\right)\right] \\
&-& \frac{\sigma}{\sqrt{2\pi}}
\left\{\exp\left[-\frac{(Jm+H_{0})^{2}}{2\sigma^{2}}\right] +
\exp\left[-\frac{(Jm-H_{0})^{2}}{2\sigma^{2}}\right]\right\}~, 
\label{16}
\end{eqnarray}

\begin{equation}
m=\frac{1}{2}{\rm erf}\left(\frac{Jm+H_{0}}{\sigma \sqrt{2}}\right) +
\frac{1}{2}{\rm erf}\left(\frac{Jm-H_{0}}{\sigma \sqrt{2}}\right)~. 
\label{17}
\end{equation}

\begin{figure}[t]
\begin{center}
\includegraphics[width=0.60\textwidth,angle=0]{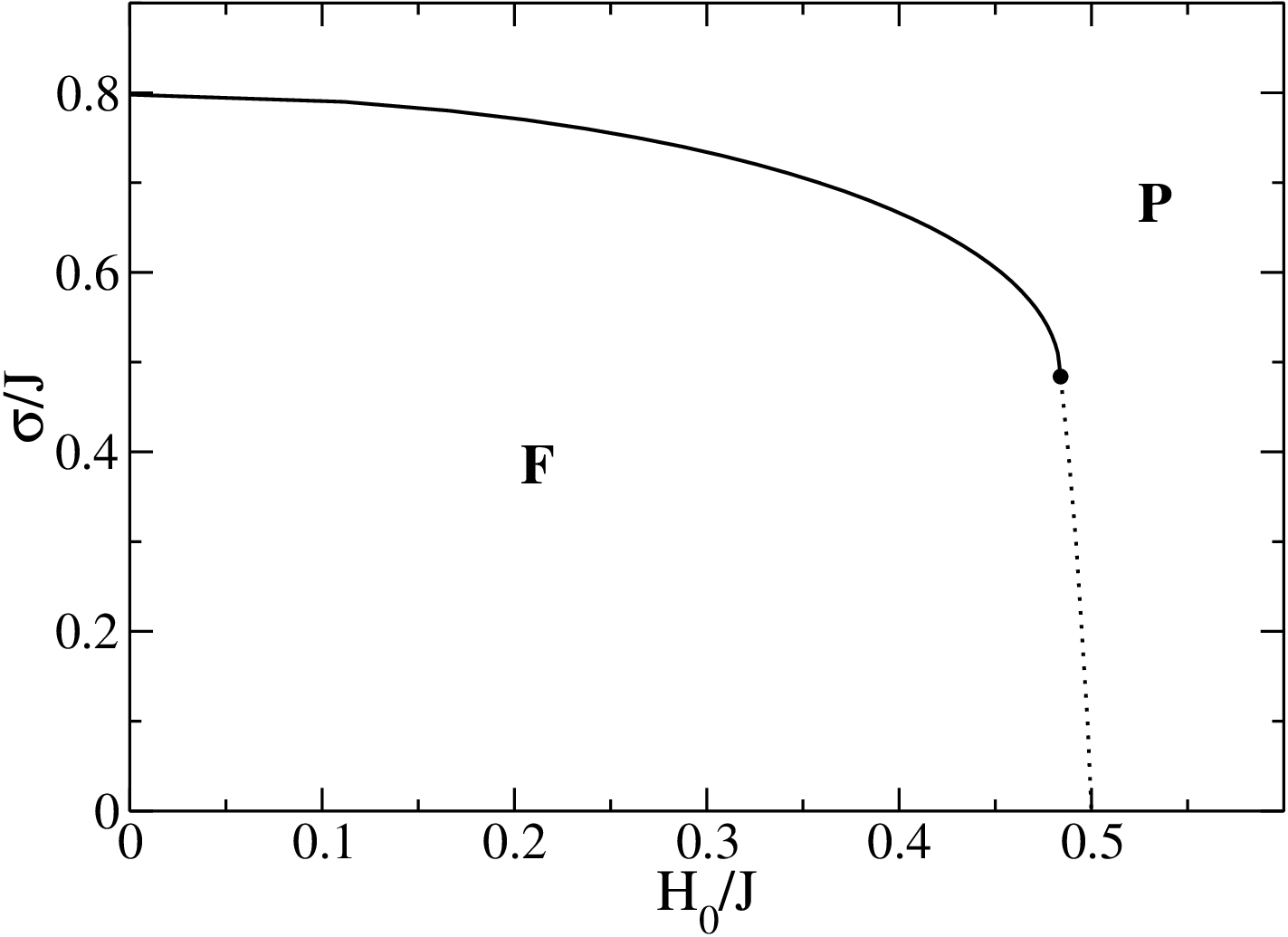}
\end{center}
\protect\caption{Phase diagram of the infinite-range-interaction 
ferromagnet in the presence of
a double Gaussian random field at zero temperature. The full
(dashed) line represents a continuous (first-order) phase transition; 
the black circle denotes a tricritical point.}
\label{fig3}
\end{figure}

\vskip \baselineskip
\noindent
A procedure similar to the one described above for finite temperatures
applies in this case, in such a way that one may expand Eq. (\ref{17}) in
powers of $m$, 

\begin{equation} \label{18}
m=am+bm^{3}+cm^{5}+O(m^{7})~, 
\end{equation}

\vskip \baselineskip
\noindent
where

\begin{eqnarray} \label{19}
a &=&
\sqrt{\frac{2}{\pi}}\left(\frac{J}{\sigma}\right)
\exp\left(-\frac{H_{0}^{2}}{2\sigma^{2}}\right)~, \\  \label{20}
b &=& \frac{1}{6}\sqrt{\frac{2}{\pi}}\left(\frac{J}{\sigma}\right)^{3}
\left\{\left(\frac{H_{0}}{\sigma}\right)^{2}-1\right\}
\exp\left(-\frac{H_{0}^{2}}{2\sigma^{2}}\right)~, \\ \label{21}
c &=& \frac{1}{120}\sqrt{\frac{2}{\pi}}\left(\frac{J}{\sigma}\right)^{5}
\left\{\left(\frac{H_{0}}{\sigma}\right)^{4} -
6\left(\frac{H_{0}}{\sigma}\right)^{2}+3\right\}
\exp\left(-\frac{H_{0}^{2}}{2\sigma^{2}}\right)~.
\end{eqnarray}

\vskip \baselineskip
\noindent
A continuous critical frontier occurs at zero temperature
for $b<0$ (i.e., $H_{0}/\sigma < 1$),
in such a way that the condition $a=1$ yields a relation involving
$H_{0}/J$ and $\sigma/J$ for this critical frontier,  

\begin{equation} \label{22}
\frac{\sigma}{J}=\sqrt{\frac{2}{\pi}}\exp\left[-\frac{1}{2}
\left(\frac{H_{0}}{J}\right)^{2}
\left(\frac{J}{\sigma}\right)^{2}\right]~.    
\end{equation}

\vskip \baselineskip
\noindent
One notices that the zero-temperature value of $H_{0}/J$ decreases, for
increasing values of $\sigma/J$, and in particular, when 
$\sigma/J = \sqrt{2/\pi}$, one gets  
$H_{0}/J=0$ as a solution. 
The critical frontier presents a tricritical point at zero-temperature
given by $b=0, \ a=1$, 

\begin{equation} \label{23}
\frac{H_{0}}{\sigma}=1~, \quad \frac{\sigma}{J}=\sqrt{\frac{2}{e\pi}}
\cong 0.4839~.  
\end{equation}

\vskip \baselineskip
\noindent
This represents the threshold value of $\sigma$ for the existence of a
tricritical point; above this value one gets that
$H_{0}/\sigma<1$, in such a way that the probability distribution of 
Eq.~(\ref{2}) presents a single maximum at the origin (cf. Fig.~1) and
there is no tricritical point, i.e., the critical frontier is completely
continuous, in agreement with previous analyses
\cite{aharony78,andelman,galambirman}.

For $b>0$, one gets a first-order critical frontier at zero-temperature,
which is associated with a tricritical point at finite temperatures. The
phase diagram of the model, at zero temperature, is presented in Fig.~3. 
Similarly to what happened in the finite-temperature phase diagram, the
first-order critical frontier changes into a continuous one, for increasing
values of $\sigma/J$, like other zero-temperature studies of
the RFIM \cite{kushauer,sethna93}.

\noindent
\section{Conclusions}

 
Summarizing, the main effects produced by an increase of $\sigma/J$ in the
present model are:  
(i) a decrease in the extension of the ferromagnetic phase; 
(ii) in the range of $\sigma/J$ for which there 
is a first-order transition line, one observes also a 
decrease in the extension of such a line; (iii) for sufficiently large
values of $\sigma/J$, the first-order transitions are transformed into
continuous ones. 
At the threshold value 
$(\sigma/J) = \sqrt{2/e\pi} \cong 0.4839$, the tricritical point where the
continuous and the first-order transition lines meet, occurs at 
zero temperature, and for greater values of $\sigma/J$, the critical
frontier is completely continuous, i.e., there is no tricritical point.
Therefore, the ratio $\sigma/J$ is directly related to the disorder in a
real system; for the case of a diluted antiferromagnet, an increase in
$\sigma/J$ should 
play a similar role as an increase in the dilution. 
 
A crossover of the phase transition from first-order to a continuous one,
due to an increase in the amount of disorder, has been observed in the
diluted antiferromagnet 
${\rm Fe_{x}Mg_{1-x}Cl_{2}}$, with $0.7 < {\rm x} < 1.0$
\cite{kushauerkleemann,kushauer}. Such an effect, that has been observed in
low -- but finite temperatures -- has been explained in
terms of zero-temperature analyses of different formulations of the RFIM 
\cite{kushauer,sethna93}. We believe that the present model is more
appropriate for a theoretical description of this effect. 
In this case, an increase in the measure of randomness in our model, 
$\sigma/J$, would be related to a decrease in the magnetic concentration
$x$, in such a way that the threshold value $(\sigma/J) = \sqrt{2/e\pi}$
would correspond to the critical value $x=0.6$, at which the first-order
transition disappears.  
In addition to that, the present model may also explain similar effects
that could possibly be  
found, at higher temperatures, on other diluted antiferromagnets.

Finaly, we argue that the double-Gaussian probability
distribuition, defined above, is suitable for an appropriate theoretical
description of the RFIM, being a better candidate for such a purpose than
the two most commonly used distributions in the literature. 

(i) In the identifications of the RFIM with diluted
antiferromagnets in the presence of a uniform magnetic field, the local
random fields are expressed in terms of quantities that vary in both sign
and magnitude \cite{fishmanaharony,cardy}; this characteristic rules out
the bimodal probability distribution from such a class of physical systems.

(ii) Although the RFIM defined in terms of a simple Gaussian probability
distribution for the fields is physically acceptable, it usually
leads to a continuous phase transition at finite temperatures, either within
mean-field \cite{aharony78,andelman,galambirman}, or standard
short-range-interaction approaches \cite{gofman,swift}. 
Such a system is not able to exhibit
first-order phase transitions and tricritical points, that may occur in some
diluted antiferromagnets \cite{belangerreview}. 

(iii) By varying appropriately the ratio $H_{0}/\sigma$ (a ratio related to
the external applied uniform field and the dilution in a real system)
in the double-Gaussian probability distribution of the present RFIM, one
may adjust the model to given physical situations, in order to
reproduce a wide variety of 
physical effects that occur in diluted antiferromagnets, like continuous
and first-order phase transitions, as well as tricritical points.

\vskip 2\baselineskip

{\large\bf Acknowledgments}

\vskip \baselineskip
\noindent
We thank Prof. Evaldo M.~F. Curado for fruitful
conversations. The partial financial supports from
CNPq and Pronex/MCT/FAPERJ (Brazilian agencies) are acknowledged. 

\vskip 2\baselineskip

\end{document}